\documentclass[onecolumn,final]{IEEEtran}

\usepackage{times}
\usepackage{amsmath}
\usepackage{amssymb}
\usepackage{amsthm}
\usepackage{color}
\usepackage{colortbl}
\usepackage{float}
\usepackage{cite}
\usepackage[caption=false, ...]{subfig}
\usepackage{verbatim}
\usepackage[pdftex]{graphicx}

\newtheorem{theorem}{Theorem}%[section]
\newtheorem{proposition}[theorem]{Proposition}%[section]
\newtheorem{corollary}[theorem]{Corollary}%[section]
\newtheorem{lemma}[theorem]{Lemma}%[section]
\newtheorem{definition}[theorem]{Definition}%[section]
\newtheorem{remark}[theorem]{Remark}%[section]

\newfont{\bbb}{msbm10 scaled 500}

\newfont{\bb}{msbm10 scaled 1100}

\newcommand{\FF}{\mbox{\bb F}}

\newcommand{\cv}{{\bf c}}

\newcommand{\fv}{{\bf f}}

\newcommand{\Cc}{{\cal C}}

\newcommand{\Fc}{{\cal F}}
\newcommand{\Gc}{{\cal G}}

\newcommand{\Jc}{{\cal J}}

\newcommand{\Mc}{{\cal M}}

\newcommand{\Pc}{{\cal P}}

\definecolor{OXO-emph}{RGB}{153,0,0}
\title{Repairable Block Failure Resilient Codes}

\author{\IEEEauthorblockN{Gokhan Calis and O.~Ozan~Koyluoglu}\\
\IEEEauthorblockA{
Department of Electrical and Computer Engineering\\
The University of Arizona\\
Email: \{gcalis, ozan\}@email.arizona.edu}
}

\begin{document}

\maketitle

%%%%%%%%%%%%%%%%%%%%%%%%%%%%%%%%%%%%%%%%%%%%%%%%%%%%%%%%%%%%%%%%%%%%%%%%%%%%%%
%%%%%%%%%%%%%%%%%%%%%%%%%%%%%%%%%%%%%%%%%%%%%%%%%%%%%%%%%%%%%%%%%%%%%%%%%%%%%%

\begin{abstract}
In large scale distributed storage systems (DSS) deployed in cloud computing, correlated failures resulting in simultaneous failure (or, unavailability) of blocks of nodes are common. In such scenarios, the stored data or a content of a failed node can only be reconstructed from the available live nodes belonging to available blocks. To analyze the resilience of the system against such block failures, this work introduces the framework of Block Failure Resilient (BFR) codes, wherein the data (e.g., file in DSS) can be decoded by reading out from a same number of codeword symbols (nodes) from each available blocks of the underlying codeword. Further, repairable BFR codes are introduced, wherein any codeword symbol in a failed block can be repaired by contacting to remaining blocks in the system. Motivated from regenerating codes, file size bounds for repairable BFR codes are derived, trade-off between per node storage and repair bandwidth is analyzed, and BFR-MSR and BFR-MBR points are derived. Explicit codes achieving these two operating points for a wide set of parameters are constructed by utilizing combinatorial designs, wherein the codewords of the underlying outer codes are distributed to BFR codeword symbols according to projective planes.
\end{abstract}

%\begin{IEEEkeywords}
%Combinatorial designs, distributed storage systems, locally repairable codes, maximum rank distance codes, regenerating codes
%\end{IEEEkeywords}

%%%%%%%%%%%%%%%%%%%%%%%%%%%%%%%%%%%%%%%%%%%%%%%%%%%%%%%%%%%%%%%%%%%%%%%%%%%%%%
%%%%%%%%%%%%%%%%%%%%%%%%%%%%%%%%%%%%%%%%%%%%%%%%%%%%%%%%%%%%%%%%%%%%%%%%%%%%%%

\section{Introduction}
Increasing demand for storing and analyzing \textit{big-data} as well as several applications of cloud computing systems require efficient cloud computing infrastructures. One inevitable nature of the storage systems is node failures. In order to provide resilience against failures, redundancy is introduced in the storage. Classical redundancy schemes range from \emph{replication} to \emph{erasure coding}. Erasure coding allows for better performance in terms of reliability and redundancy compared to replication, however repair bandwidth in reconstructing a failed node is higher. Regenerating codes are proposed to overcome this problem in the seminal work of Dimakis et al. \cite{Dimakis:Network10}. In such a model of distributed storage systems (DSS), the file $\Mc$ is encoded to $n$ nodes such that any $k\leq n$ nodes (each with $\alpha$ symbols) allow for reconstructing the file and any $d\geq k$ nodes (with $\beta\leq \alpha$ symbols from each) reconstruct a failed node with a repair bandwidth $\gamma=d\beta$. The trade-off between per node storage ($\alpha$) and repair bandwidth ($\gamma$) is characterized and two ends of the trade-off are named as minimum storage regenerating (MSR) and minimum bandwidth regenerating (MBR) points \cite{Dimakis:Network10}. Several explicit codes have been proposed to achieve these points recently \cite{Tamo:Zigzag13,Rashmi:Optimal11,Dimakis:Survey11}. Another metric for an efficient repair is repair degree $d$, and regenerating codes necessarily have $d\geq k$. Codes with locality and locally repairable codes with regeneration properties~\cite{Gopalan:Locality12,Papailiopoulos:Locally12,Oggier:Self11,Rawat:Optimal14,Kamath:Codes12,Kamath:Explicit13} allow for a small repair degree, wherein failed nodes are reconstructed via local connections. Instances of such codes are recently considered in DSS~\cite{Sathiamoorthy:XORing13,Huang:Erasure12}. 

In large-scale distributed storage systems (such as GFS \cite{Ghemawat:Google03}),  \textit{correlated failures} are unavoidable. As analyzed in~\cite{Ford:Availability10}, these simultaneous failures of multiple nodes affect the performance of computing systems severely. The analysis in~\cite{Ford:Availability10} further shows that these correlated failures arise due to \textit{failure domains}. For example, nodes connected to the same power source or nodes belonging to the same rack exhibit these failure bursts. The unavailability periods are transient, and largest failure bursts almost always have significant rack-correlation. In order to overcome from failures having such patterns, a different approach is needed.

In this paper, we develop a framework to analyze resilience against block failures in DSS with node repair efficiencies. We consider a DSS with a single failure domain, where nodes belonging to the same failure group constitute a block of the codeword. We introduce block failure resilient (BFR) codes, which allow for data collection from any $b_c= b-\rho$ blocks, where $b$ is the number of blocks, and $\rho$ is the resilience parameter of the code. Considering a load-balancing among blocks, a same number of nodes are contacted within these $b_c$ blocks. (A total of $k=k_cb_c$ nodes and downloading $\alpha$ - i.e., all - symbols from each.) This constitutes data collection property of BFR codes. ($\rho=0$ case can be considered as a special case of batch codes introduced in ~\cite{Ishai:Batch04}.) Then, we introduce repairability in BFR codes, where any node of a failed block can be reconstructed from any $d_r$ of any remaining $b_r\leq b-1$ blocks. (A total of $d=d_rb_r$ nodes and downloading $\beta$ symbols from each.) As introduced in~\cite{Dimakis:Network10}, we utilize graph expansion of DSS employing these repairable codes, and derive file size bounds and characterize BFR-MBR and BFR-MSR points. (We note that the blocks in our model can be used to model racks in DSS. Such a model is related to the work \cite{Gaston:realistic13} which differentiates between within-rack communication and cross-rack communication. Our focus here would correspond to the case where within rack communication is much higher than the cross-rack communication, as no nodes from the failed rack can be contacted to regenerate a node.) We construct explicit codes achieving these points for a wide set of parameters.  For a system with $b=2$ blocks case, we show that achieving both MSR and MBR properties simultaneously is asymptotically possible. (This is somewhat similar to the property of Twin codes \cite{Rashmi:Enabling11}, but here the data collection property is different.) Then, for a system with $b\geq 3$ blocks case, we consider utilizing multiple codewords, which are placed into DSS via a combinatorial design based codeword placement algorithm. We show this technique establishes optimal codes for a wide set of parameter ranges.

The paper is organized as follows. Section II introduces model and preliminaries. Section III is devoted to the analysis of file size bounds. Code constructions are provided in Section IV. Section V includes extensions and concluding remarks.

%%%%%%%%%%%%%%%%%%%%%%%%%%%%%%%%%%%%%%%%%%%%%%%%%%%%%%%%%%%%%%%%%%%%%%%%%%%%%%
%%%%%%%%%%%%%%%%%%%%%%%%%%%%%%%%%%%%%%%%%%%%%%%%%%%%%%%%%%%%%%%%%%%%%%%%%%%%%%

\section{Background and Preliminaries}
\label{sec:Background}
\subsection{Block failure resilient codes and repairability}
Consider a code $\Cc$ which maps $\Mc$ symbols (over $\FF_q$) in $\fv$ (file) to length $n$ codewords (nodes) $\cv=(c_1,\cdots,c_n)$ with $c_i\in \FF_q^\alpha$ for $i=1,\cdots,n$. These codewords are distributed into $b$ blocks each with block capacity $c=\frac{b}{n}$ nodes per block. We have the following definition.

\begin{definition}[Block Failure Resilient (BFR) Codes]
An $(n,b,\Mc,k,\rho,\alpha)$ block failure resilient (BFR) code encodes $\Mc$ elements in $\FF_q$ ($\fv$) to $n$ codeword symbols (each in $\FF_q^\alpha$) that are grouped into $b$ blocks such that $\fv$ can be decoded by accessing to any $\frac{k}{b-\rho}$ nodes of from each of the  $b-\rho$ blocks.
\end{definition}

We remark that, in the above, $\rho$ represents the resilience parameter of the BFR code, i.e., the code can tolerate $\rho$ block erasures. Due to this data collection (file decoding) property of the code, we denote the number of blocks accessed as $b_c=b-\rho$ and number of nodes accessed per block as $k_c=\frac{k}{b_c}$. Noting that $k_c\leq c$ should be satisfied, we differentiate between \emph{partial} block access, $k_c<c$, and \emph{full} block access $k_c=c$. Throughout the paper, we assume $n|b$. i.e., $c$ is integer, and $(b-\rho)|k$, i.e., $k_c$ is integer.

%Scalar BFR codes (i.e., $\alpha=1$ in the definition above) with $\rho=0$ are special cases of batch codes introduced in ~\cite{Ishai:Batch04}. In particular, an $(n',N',k_c,m',t')$ batch code encodes an $n'$-length string to $m'$ buckets (of total length $N'$) such that any length $k_c$-tuple of the string can be decoded by accessing to at most $t'$ symbols per bucket. Scalar BFR codes with $b_c=b$ correspond to accessing all buckets as in batch codes but with recovering all of the underlying string, i.e., $k_c=n'$ case.

Remarkably, any MDS array code \cite{McWilliams:Theory77} can be utilized as BFR codes for the full access case. In fact, such an approach will be optimal in terms of minimum distance, and therefore for resilience $\rho$. However, for $k_c<c$, MDS array codes may not result in an optimal code. Constructing optimal BFR codes in terms of the trade-off between resilience $\rho$ and code rate $\frac{\Mc}{n\alpha}$ will be studied elsewhere. In this work, we focus on repairable BFR codes, as defined in the following.

\begin{definition}[Block Failure Resilient Regenerating Codes (BFR-RC)]
An $(n,b,\Mc,k,\rho,\alpha,d,\sigma,\beta)$ block failure resilient regenerating code (BFR-RC) is an $(n,b,\Mc,k,\rho,\alpha)$ BFR code (data collection property) with the following repair property: Any node of a failed block can be reconstructed by accessing to any $d_r=\frac{d}{b-\sigma}$ nodes of any $b_r=b-\sigma$ blocks and downloading $\beta$ symbols from each of these $d=b_rd_r$ nodes.
\end{definition}

We assume $(b-\rho)|d$, i.e., $d_r$ is integer. (Note that $d_r$ should necessarily satisfy $\frac{d}{b-\sigma}=d_r\leq c=\frac{n}{b}$ in our model.) We consider the trade-off between the \emph{repair bandwidth} $\gamma=d\beta$ and \emph{per node storage} $\alpha$ similar to the seminal work ~\cite{Dimakis:Network10}. In particular, we define $\alpha_{\textrm{BFR-MSR}}=\frac{\Mc}{k}$ as the minimum per node storage and $\gamma_{\textrm{BFR-MBR}}=\alpha$ as the minimum repair bandwidth for an $(n,b,\Mc,k,\rho,\alpha,d,\sigma,\beta)$ BFR-RC.
When deriving this trade-off, we focus on systems having $d_r=\frac{d}{b-\sigma}\geq k_c=\frac{k}{b-\rho}$, i.e., data collection process contacts to less number of nodes per block as compared to symbol regeneration. (We note that, similar to regenerating codes, without loss of generality, one should only consider systems that satisfy $d\geq k$, i.e., $d_r(b-\sigma)\geq k_c (b-\rho)$. Therefore, our $d_r\geq k_c$ assumption can be made without loss of generality for systems having $\rho\leq \sigma$.)

\subsection{Information flow graph}
The operation of a DSS employing such codes can be modeled by a multicasting scenario over an information flow graph \cite{Dimakis:Network10}, which has three types of nodes: 1) Source node ($S$): Contains original file $\fv$. 2) Storage nodes, each represented as $x_i$ with two sub-nodes($(x^{\rm in}_i,x^{\rm out}_i)$), where $x^{\rm in}$ is the sub-node having the connections from the live nodes, and $x^{\rm out}$ is the storage sub-node, which stores the data and is contacted for node repair or data collection (edges between each $x^{\rm in}_i$ and $x^{\rm out}_i)$ has $\alpha$-link capacity). 3) Data collector ($\rm{DC}$) which contacts $x^{\rm{out}}$ sub-nodes of $k$ live nodes (with edges each having $\infty$-link capacity). (As described above, for BFR codes these $k$ nodes can be any $\frac{k}{b-\rho}$ nodes from each of the $b-\rho$ blocks.) Then, for a given graph $\Gc$ and DCs $\rm{DC}_i$, the file size can be bounded using the max flow-min cut theorem for multicasting utilized in network coding~\cite{Ho:random06,Dimakis:Network10}.
\begin{lemma}[Max flow-min cut theorem for multicasting] \label{lma:MFMCforMulticast}
$$\Mc \leq \min_{\Gc} \min_{\rm{DC}_i} \rm{max flow}(S \to \rm{DC}_i,\Gc),$$
where $\rm{flow}(S \to \rm{DC}_i,\Gc)$ represents the flow from the source
node $S$ to $\rm{DC}_i$ over the graph $\Gc$.
\end{lemma}
Therefore, $\mathcal{M}$ symbol long file can be delivered to a DC, only if the min cut is at least $\mathcal{M}$. In the next section, similar to Dimakis et al., \cite{Dimakis:Network10}, we consider $k$ successive node failures and evaluate the min-cut over possible graphs, and obtain a file size bound for a DSS operating with BFR-RC.

\subsection{Block designs and projective planes}

%We first provide the definition of balanced incomplete block designs (BIBDs), which are probably one of the most-studied types of block designs \cite{Stinson:Combinatorial04}.
We first provide the definition of balanced incomplete block designs (BIBDs) \cite{Stinson:Combinatorial04}.
\begin{definition}[Balanced incomplete block design]
A $(v,\kappa,\lambda)$-BIBD has $v$ points distributed into blocks of size $\kappa$ such that any pair of points are contained in $\lambda$ blocks.
\end{definition}

%A $(v,\kappa,\lambda)$-BIBD then have the following properties \cite{Stinson:Combinatorial04}.
\begin{corollary}\label{thm:BIBDcorollary}
For a $(v,\kappa,\lambda)$-BIBD,
\begin{itemize}
\item Every point occurs in $r=\frac{\lambda (v-1)}{\kappa-1}$ blocks.
\item The design has exactly $b=\frac{vr}{\kappa}=\frac{\lambda(v^2-v)}{\kappa^2-\kappa}$ blocks.
\end{itemize}
\end{corollary}

In the achievable schemes of this work, we utilize a special class of block designs that are called projective planes.
\begin{definition}
A $(v=p^2+p+1,\kappa=p+1,\lambda=1)$-BIBD with $p\geq 2$ is called a projective plane of order $p$.
\end{definition}

Projective planes have the property that every pair of blocks intersect at a unique point (as $\lambda=1$). In addition, due to Corollary~\ref{thm:BIBDcorollary}, in projective planes, every point occurs in $r=p+1$ blocks, and there are $b=v=p^2+p+1$ blocks.

%%%%%%%%%%%%%%%%%%%%%%%%%%%%%%%%%%%%%%%%%%%%%%%%%%%%%%%%%%%%%%%%%%%%%%%%%%%%%%
%%%%%%%%%%%%%%%%%%%%%%%%%%%%%%%%%%%%%%%%%%%%%%%%%%%%%%%%%%%%%%%%%%%%%%%%%%%%%%

\section{File Size Bound for Repairable BFR Codes}
\label{sec:FileSize}

Information flow graph analysis, similar to that of considered in \cite{Dimakis:Network10}, can be performed to obtain file size bounds for repairable BFR codes. In this paper, we focus on the case $\sigma=1$, i.e., regeneration of a node in a failed block is performed by contacting to all remaining live blocks. In the following, we first analyze $\rho=0$ case, i.e., data collector connects all the blocks to reconstruct the data.

\subsection{$\rho=0$, $b=2$ case}

\begin{figure}[t]
 \centering
 \includegraphics[width=0.6\columnwidth]{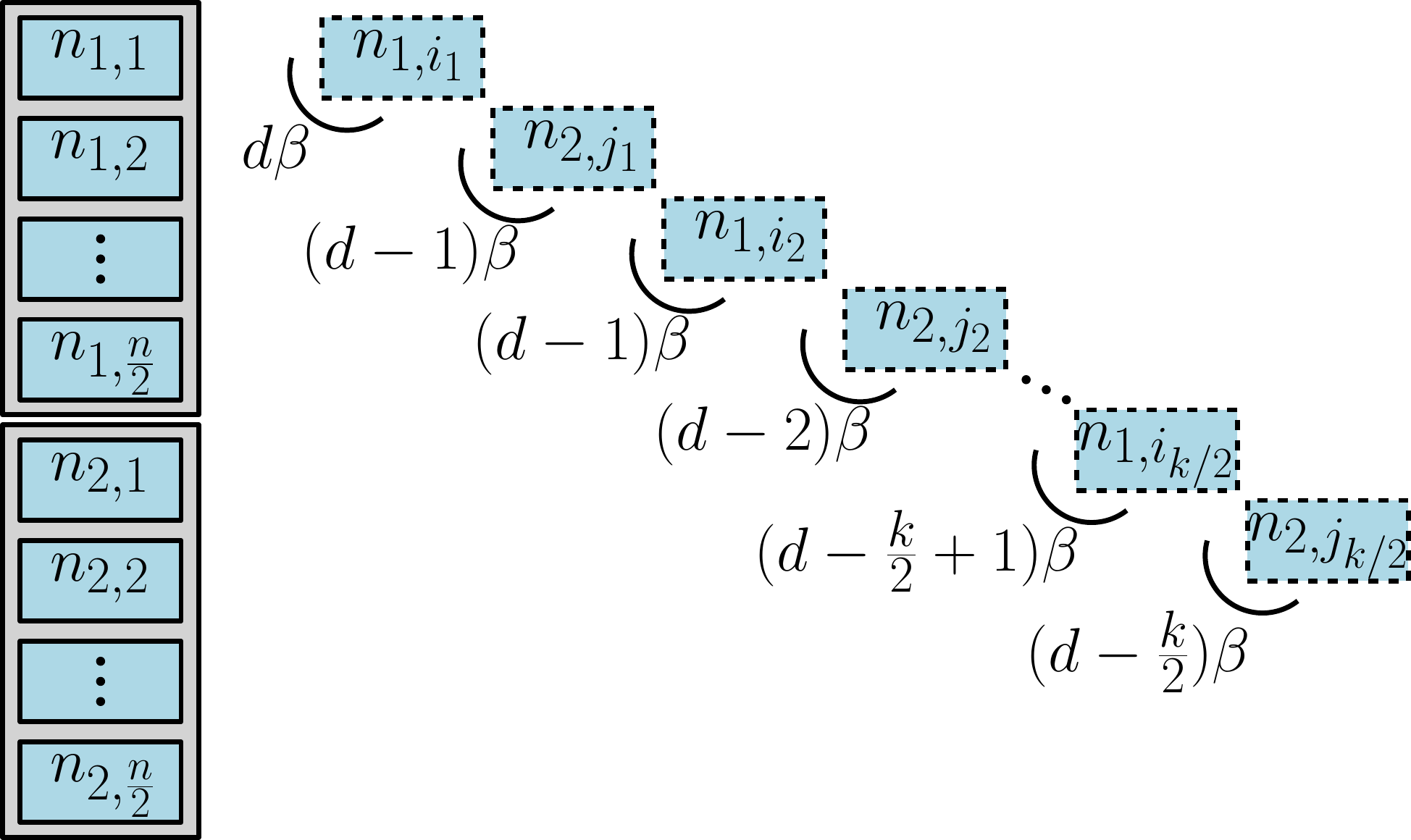}
 \caption{Repair process for $b=2$ (two blocks) case.}
\label{fig:Two-block}
\end{figure}

Consider $b=2$-block case as in Fig.~\ref{fig:Two-block} and assume $2|k$. From Lemma~\ref{lma:MFMCforMulticast}, the file size $\Mc$ can be upper bounded with the repair procedure shown in Fig.~\ref{fig:Two-block}, which displays one of the ``minimum-cut'' scenarios, wherein any two consecutive node failures belong to different blocks. Assuming $k$ is even and $d\geq \frac{k}{2}$, 
\begin{equation}
\Mc \leq \sum_{i=0}^{\frac{k}{2}-1}\min(\alpha,(d-i)\beta) + \sum_{i=1}^{\frac{k}{2}}\min(\alpha,(d-i)\beta).
\label{eq:min-cut_two}
\end{equation}
Achieving this upper bound \eqref{eq:min-cut_two} with equality would yield maximum possible file size. One particular instance is shown in Fig.~\ref{fig:Two-block}, and we note that the order of failed nodes does not matter as the sum of the cut would be the same with different order of failures as long as we consider connection from data collector to $\frac{k}{2}$ repaired nodes from each block.

For MSR point, $\alpha=\alpha_{\textrm{BFR-MSR}}=\frac{\Mc}{k}$. In the bound \eqref{eq:min-cut_two}, we then have $\alpha_{\textrm{BFR-MSR}} \leq (d-\frac{k}{2})\beta_{\textrm{BFR-MSR}}$. Achieving equality would give the minimum repair bandwidth for the MSR case. Hence, BFR-MSR point is given by
\begin{equation}
(\alpha_{\textrm{BFR-MSR}},\gamma_{\textrm{BFR-MSR}}) = (\frac{\Mc}{k},\frac{2\Mc d}{2kd-k^2}).
\label{eq:min-cut_two-MSR-values}
\end{equation}

%BFR-MBR codes, on the other hand, have the property that $d\beta=\alpha$ with minimum possible $d\beta$ while achieving the equality in \eqref{eq:min-cut_two}. Then,
%\begin{equation}
%\Mc = \sum_{i=0}^{k/2-1}(d-i)\beta + \sum_{i=1}^{k/2}(d-i)\beta.
%\label{eq:min-cut_two-MBR}
%\end{equation}
BFR-MBR codes, on the other hand, have the property that $d\beta=\alpha$ with minimum possible $d\beta$ while achieving the equality in \eqref{eq:min-cut_two}. Inserting $d\beta=\alpha$ in \eqref{eq:min-cut_two}, we obtain that
%Then, from (\ref{eq:min-cut_two-MBR}), we obtain that
\begin{equation}
(\alpha_{\textrm{BFR-MBR}},\gamma_{\textrm{BFR-MBR}}) = (\frac{4\Mc d}{4dk-k^2},\frac{4\Mc d}{4dk-k^2}).
\label{eq:min-cut_two-MBR-values}
\end{equation} 
Same analysis can be done for odd values of $k$ as well, %resulting in the following.
\begin{equation}
(\alpha_{\textrm{BFR-MSR}},\gamma_{\textrm{BFR-MSR}}) =
\begin{cases}
	(\frac{M}{k},\frac{2Md}{2kd-k^2-k}), \textrm{ if $k$ is odd} \\
    (\frac{M}{k},\frac{2Md}{2kd-k^2}), \textrm{ o.w.}
\end{cases}
\label{eq:MSR_cases}
\end{equation}
\begin{equation}
(\alpha_{\textrm{BFR-MBR}},\gamma_{\textrm{BFR-MBR}}) =
\begin{cases}
	(\frac{4Md}{4dk-k^2+1},\frac{4Md}{4dk-k^2+1}), \textrm{ if $k$ is odd} \\
    (\frac{4Md}{4dk-k^2},\frac{4Md}{4dk-k^2}), \textrm{ o.w.}
\end{cases}
\label{eq:MBR_cases}
\end{equation}
%One important observation from (\ref{eq:MSR_cases}) is that $\alpha_{\textrm{BFR-MSR}}=\alpha_{\textrm{MSR}}$ and $\gamma_{\textrm{BFR-MSR}}=\gamma_\textrm{MBR}$, i.e., the per node storage of MSR point and repair bandwidth of MBR point of regenerating codes \cite{Dimakis:Network10} are simultaneously achievable. We provide the generalization of these bounds to $b \geq 2$ case in the following.
Here, we compare $\gamma_{\textrm{BFR-MSR}}$ and $\gamma_{\textrm{MBR}}$. We have $\gamma^{\textrm{k-odd}}_{\textrm{BFR-MSR}} \geq \gamma^{\textrm{k-even}}_{\textrm{BFR-MSR}} \geq \gamma_{\textrm{MBR}}=\frac{2\Mc d}{k(2d-k+1)}$, and, if we have $2d-k \gg 1$, then $\gamma^{\textrm{k-odd}}_{\textrm{BFR-MSR}} \approx \gamma^{\textrm{k-even}}_{\textrm{BFR-MSR}} \approx \gamma_{\textrm{MBR}}$. This implies that BFR-MSR codes with $b=2$ achieves repair bandwidth of MBR and per-node storage of MSR codes simultaneously for systems with $d \gg 1$. We provide the generalization of these bounds to $b \geq 2$ case in the following.

\subsection{$\rho=0$, $b\geq2$ case}

The same steps described above can be used to derive the file size bound for $b$-blocks.
%Accordingly, the optimal file size for a DSS employing a repairable BFR code is as follows.

\begin{lemma}
The optimal file size is given by
\begin{align}
\begin{split}
\Mc  =  & \sum_{i=0}^{\frac{k}{b}-1}\min(\alpha,(d-(b-1)i)\beta) \\
& + \sum_{i=0}^{\frac{k}{b}-1}\min(\alpha,(d-1-(b-1)i)\beta) + \ldots \\
& + \sum_{i=0}^{\frac{k}{b}-1}\min(\alpha,(d-(b-1)-(b-1)i)\beta).
\end{split}
\end{align}
\end{lemma}  

\begin{proposition}
BFR-MSR and BFR-MBR points are as follows,
\begin{equation}
(\alpha_{\textrm{BFR-MSR}},\gamma_{\textrm{BFR-MSR}}) = \left(\frac{\Mc}{k},\frac{\Mc d}{kd-\frac{k^2(b-1)}{b}}\right)
\label{eq:min-cut_b-MSR-values}
\end{equation}

\begin{equation}
(\alpha_{\textrm{BFR-MBR}},\gamma_{\textrm{BFR-MBR}}) = \left(\frac{\Mc d}{kd-\frac{k^2(b-1)}{2b}},\frac{\Mc d}{kd-\frac{k^2(b-1)}{2b}}\right)
\label{eq:min-cut_b-MBR-values}
\end{equation}
\end{proposition}

We observe that $\gamma_{\textrm{BFR-MSR}} \leq \gamma_{\textrm{MSR}}=\frac{\Mc d}{k(d-k+1)}$ for $b \leq k$, which is the case here as $b \mid k$. Also, we have $\frac{\gamma_{\textrm{BFR-MSR}}}{\gamma_{\textrm{MBR}}} = \frac{d-\frac{k-1}{2}}{d-k\frac{b-1}{b}} \geq 1$ when $b \geq \frac{2k}{k+1}$ which is always true. Hence, $\gamma_{\textrm{BFR-MSR}}$ is between $\gamma_{\textrm{MSR}}$ and $\gamma_{\textrm{MBR}}$.  

\subsection{$\rho>0$ case}

If we restrict data collector to connect $b_c<b$ blocks (i.e., $\rho>0$), but keep the repair process same as before, the above analysis follows and corresponding MSR and MBR points are given by replacing $b$ in \eqref{eq:min-cut_b-MSR-values} and \eqref{eq:min-cut_b-MBR-values} with $b_{c}=b-\rho$ - for systems satisfying $d_r\geq k_c$. (This follows as the repair from these $\rho$ blocks will not contribute to the cut between the source $S$ and DC.)

%%%%%%%%%%%%%%%%%%%%%%%%%%%%%%%%%%%%%%%%%%%%%%%%%%%%%%%%%%%%%%%%%%%%%%%%%%%%%%
%%%%%%%%%%%%%%%%%%%%%%%%%%%%%%%%%%%%%%%%%%%%%%%%%%%%%%%%%%%%%%%%%%%%%%%%%%%%%%

\section{BFR-MSR and BFR-MBR Code Constructions}
\label{sec:CodeConst}

\subsection{Transpose code for b=2 case}

\begin{figure}[t]
 \centering
 \includegraphics[width=0.60\columnwidth]{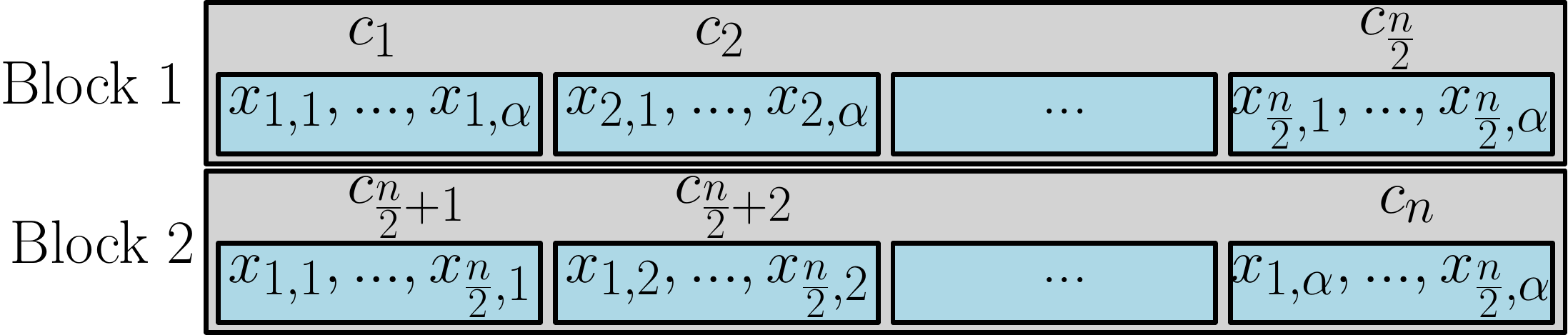}
 \caption{Transpose code is a two-block BFR-MBR code.}
\label{fig:Transpose}
\end{figure}

One instance of BFR codes is given in the Fig.~\ref{fig:Transpose}. We set $\alpha=d=\frac{n}{2}$, and store the transpose of the first block's symbols in the second block. The repair of a failed node $i$ in the first block can be performed by connecting all the nodes in the second block and downloading only $1$ symbol from each node. That is, $d\beta=\alpha$. Further, we set $\Mc=kd-(\frac{k}{2})^2$, and use an $[N=\alpha^2,K=\Mc]$ MDS code to encode file $\fv$ into symbols denoted with $x_{i,j}$, $i,j=1,...,\alpha$. BFR data collection property allows for reconstructing the file, as connecting any $\frac{k}{2}$ nodes from each block assures at least $K$ distinct symbols. This code is a BFR-MBR code for $\beta=1$ (scalar code), as the optimal file size in \eqref{eq:min-cut_two-MBR-values}, i.e., $\Mc=kd-(\frac{k}{2})^2$, is achieved with $d\beta=\alpha$. A similar code to this construction is Twin codes introduced in \cite{Rashmi:Enabling11}, where the nodes are split into two types and a failed node of a a given type is regenerated by connecting to nodes only in the other type. However, Twin codes, as opposed to our model, do not have balanced node connection for data collection. In particular, DC connects to only (a subset of $k$ nodes from) a single type. On the other hand, BFR codes, for $b=2$ case, connects to $\frac{k}{2}$ nodes from each block.

\subsection{Block design based regenerating code symbol placement}

\begin{figure}[t]
 \centering
 \includegraphics[width=0.6\columnwidth]{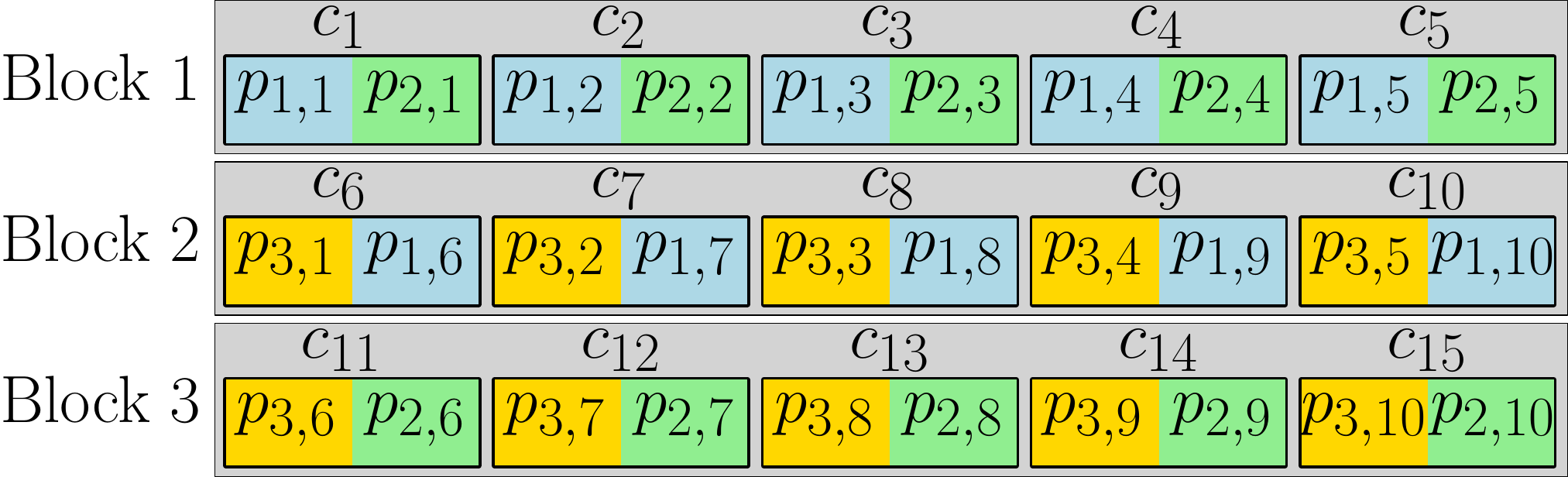}
 \caption{Three-block BFR-RC via projective plane symbol placement.}
\label{fig:3-block BFR-RC}
\end{figure}

Consider that the file $\Fc$ of size $\Mc$ contains 3 sub-files $\Fc_{1}$, $\Fc_{2}$ and $\Fc_{3}$ each of size $\tilde{\Mc}$. We encode these sub-files with $[\tilde{n}=10,\tilde{k}=4,\tilde{d}=5,\tilde{\alpha},\tilde{\beta}]$ regenerating code $\tilde{\Cc}$, represent the resulting symbols with $\Pc_{1}=p_{1,1:\tilde{n}}$ for $\Fc_1$, $\Pc_{2}=p_{2,1:\tilde{n}}$ for $\Fc_2$, and $\Pc_{3}=p{_{3,1:\tilde{n}}}$ for $\Fc_3$. These symbols are grouped in a specific way placed into nodes within blocks as represented in Fig.~\ref{fig:3-block BFR-RC}, where each node contains two symbols each coming from two of the different sets $\Pc_{1},\Pc_{2},\Pc_{3}$. We set the sub-code $\tilde{\Cc}$ parameters as $[\Mc=3\tilde{\Mc}, k=\frac{3}{2}\tilde{k}, d=2\tilde{d}, \alpha=2\tilde{\alpha}, \beta=\tilde{\beta}]$.

Assume Block 1 is unavailable and its first node, which contains codeword $c_{1}$, has to be reconstructed. Due to underlying regenerating code, contacting $5$ nodes of Block 2 and accessing to $p_{1,6:10}$ repairs $p_{1,1}$. Similarly, $p_{2,1}$ can be reconstructed from Block 3. Any node failures can be handled similarly, by connecting to remaining 2 blocks and repairing each symbol of lost node by connecting $\tilde{d}$ nodes in a block. As we have $k=6$, DC, connecting to 2 nodes from each block, obtains $12$ symbols which has 4 different symbols from each of $\Pc_{1}$, $\Pc_{2}$ and $\Pc_{3}$. As the embedded regenerating code has $\tilde{k}=4$, all $3$ sub-files can be recovered. 

We generalize the BFR-RC construction above utilizing projective planes. First, the file $\fv$ of size $\Mc$ is partitioned into $v$ parts, $\Mc_{1}$, $\Mc_{2}$,...,$\Mc_{v}$. Each part, of size $\tilde{\Mc}$, then encoded using $[\tilde{n},\tilde{k},\tilde{d},\tilde{\alpha},\tilde{\beta}]$ regenerating code $\tilde{\Cc}$. We represent the resulting symbols with $\Pc_{i}=p_{i,1:\tilde{n}}$ for $i=1,\cdots, v$. We then consider index of each part as a point in a $(v=p^2+p+1,\kappa=p+1,\lambda=1)$ projective plane. (Indices of symbol sets $\Pc_\Jc$ and points $\Jc$ of projective plane are used interchangeably in the following.) We perform the placement of each point in the system using this projective plane mapping. (The setup in Fig. \ref{fig:3-block BFR-RC} can be considered as a toy model. Although the combinatorial design with blocks given by $\{p_1,p_2\},\{p_1,p_3\},\{p_2,p_3\}$ has projective plane properties, it is not considered as an instance of a projective plane.) In this placement, total of $\tilde{n}$ nodes from each partition $\Pc_{i}$ are distributed to $r$ blocks evenly, each block contains $\frac{\tilde{n}}{r}$ nodes where each node stores $\alpha=\kappa\tilde{\alpha}$ symbols. Note that blocks of projective plane give the indices of parts $\Pc_i$ stored in the nodes of the corresponding block in DSS. That is, all nodes in a block stores symbols from unique subset of $\Pc=\{\Pc_1,\cdots,\Pc_v\}$ of size $\kappa$. Overall, the system can store a file of size $\Mc=v\tilde{\Mc}$ with $b=v$ blocks. We set the sub-code $\tilde{\Cc}$ parameters as %follows
\begin{equation}
M=v\tilde{M}, k=\frac{b}{r}\tilde{k}, d=\kappa\tilde{d}, \alpha=\kappa\tilde{\alpha}, \beta=\tilde{\beta} 
\label{eq:assignment-b}
\end{equation}
where we choose parameters to satisfy $r-1 \mid \tilde{d}$, $r \mid \tilde{n}$ and $r \mid \tilde{k}$.

\emph{Node Repair:} Consider that one of the nodes in a block is to be repaired. Note that the failed node contains $\kappa$ symbols, each coming from a distinct subfile's regenerating codeword. Using projective planes' property that any $2$ blocks has only $1$ point in common, any remaining block can help for in the regeneration of $1$ symbol of the failed node. Furthermore, as any point has a repetition degree of $r$, one can connect to $r-1$ blocks, $\frac{\tilde{d}}{r-1}$ nodes per block, to repair one symbol of a failed node. Combining these two, node regeneration is performed by connecting $(r-1)\kappa$ blocks. Substituting $\kappa=p+1$ and $r=p+1$, connecting to $p^{2}+p=b-1$ blocks allows for reconstructing any node of a failed block.

\emph{Data Collection:} DC, connects $\frac{\tilde{k}}{r}$ nodes per block from all $b_{c}=b$ blocks, i.e., a total of $k=\frac{b}{r}\tilde{k}$ nodes each having encoded symbols of $\kappa$ subfiles. These total of $v\tilde{k}$ symbols include $\tilde{k}$ symbols from each subfile, from which all subfiles, hence the file $\fv$, can be decoded.

\subsubsection{BFR-MSR}
To construct a BFR-MSR code, we set each subcode $\tilde{\Cc}$ as an MSR code, which has
\begin{equation}
\tilde{\alpha}=\frac{\tilde{\Mc}}{\tilde{k}}, \tilde{d}\tilde{\beta}=\frac{\tilde{\Mc}\tilde{d}}{\tilde{k}(\tilde{d}-\tilde{k}+1)}.
\end{equation}

This, together with \eqref{eq:assignment-b}, results in the following parameters of our BFR-MSR construction
\begin{equation}
\alpha=\tilde{\alpha}\kappa=\frac{\Mc}{k}, d\beta=\kappa \tilde{d}\tilde{\beta}=\frac{\Mc d}{k(d-\frac{k(p+1)^{2}}{p^2+p+1}+p+1)}.
\label{eq:bfr-msr}
\end{equation}

%From \eqref{eq:bfr-msr}, we observe that this code achieves the MSR point for $\tilde{k}=p+1$. 
We remark that if we utilize ZigZag codes\cite{Tamo:Zigzag13} as the sub-code $\tilde{\Cc}$ above, we have
$[\tilde{n},\tilde{k},\tilde{d}=\tilde{n}-1,\tilde{\alpha}=\tilde{r}^{\tilde{k}-1},\tilde{\beta}=\tilde{r}^{\tilde{k}-2}, \tilde{r}=\tilde{n}-\tilde{k}]$, and having $\tilde{d}=\tilde{n}-1$ requires connecting to $1$ node per block for repairs in our block model. On the other hand, product matrix MSR codes \cite{Rashmi:Optimal11} can be used as the sub-code $\tilde{\Cc}$ for any $\tilde{d} \geq 2\tilde{k}-2$, for which we do not necessarily have $\frac{\tilde{d}}{r-1}=1$. We observe from \eqref{eq:min-cut_b-MSR-values} and \eqref{eq:bfr-msr} that MSR point is achieved for $\tilde{k}=p+1$, meaning $k=b$. 

\subsubsection{BFR-MBR}

To construct a BFR-MBR code, we set each subcode $\tilde{\Cc}$ as a product matrix MBR code \cite{Rashmi:Optimal11}, which has
\begin{equation}
\tilde{\alpha}=\tilde{d}\tilde{\beta}= \frac{2\tilde{\Mc}\tilde{d}}{\tilde{k}(2\tilde{d}-\tilde{k}+1)}.
\end{equation}

This, together with \eqref{eq:assignment-b}, results in the following parameters of our BFR-MSR construction
\begin{equation}
\alpha=d\beta=\frac{2\Mc d}{k(2d-\frac{k(p+1)^2}{p^2+p+1}+p+1)}.
\label{eq:bfr-mbr}
\end{equation}

From \eqref{eq:min-cut_b-MBR-values} and \eqref{eq:bfr-mbr}, MBR point is achieved for $\tilde{k}=p+1$.

%%%%%%%%%%%%%%%%%%%%%%%%%%%%%%%%%%%%%%%%%%%%%%%%%%%%%%%%%%%%%%%%%%%%%%%%%%%%%%
%%%%%%%%%%%%%%%%%%%%%%%%%%%%%%%%%%%%%%%%%%%%%%%%%%%%%%%%%%%%%%%%%%%%%%%%%%%%%%

\section{Extensions and concluding remarks}
\label{sec:Discussion}

\subsection{$\rho>0$ case}

In the above, we considered the cases where DC connects all $b$ blocks in file reconstruction. In order to support $b_{c}<b$, we consider employing Gabidulin codes \cite{Gabidulin:Theory85} as an outer code similar to the constructions provided in \cite{Rawat:Optimal14,Kamath:Explicit13}. We briefly discuss our approach here. Detailed results will be provided elsewhere. $[N,K,D=N-K+1]_{q^{m}}$ Gabidulin code $C^{Gab}$, $m\geq N$, has a codeword $( f(g_{1}),f(g_{2}),...,f(g_{N})) \in \FF_{q^{m}}^N$, where $f(x)$ is a linearized polynomial over $\FF_{q^{m}}$ of $q$-degree $K-1$ with $K$ message symbols as its coefficients and $g_{1},g_{2},...,g_{N} \in \FF_{q^{m}}$ are linearly independent over $F_{q}$ \cite{Gabidulin:Theory85}.

\begin{remark}
Given evaluations of $f(\cdot)$ at any $K$ linearly independent (over $\FF_{q}$) points in $\FF_{q^{m}}$, one can reconstruct the message vector.
\end{remark}

Here, before partitioning the message into $v$ parts, we encode the file with a Gabidulin code first, then partition the resulting codeword into $v$ parts and follow remaining steps as before. With this approach, decoding the message at DC follows by obtaining at least $K$ independent evaluations from $k$ nodes, $k_c=\frac{k}{b_{c}}$ nodes per block from a total of $b_{c}=b-\rho$ blocks. As considered in \cite{Rawat:Optimal14,Kamath:Explicit13}, the number of such evaluations can be derived from the rank accumulation profile of the inherent MSR/MBR codes $\tilde{\Cc}$ as in the following
\begin{equation}
\tilde{a_{j}} =
\begin{cases}
    \tilde{\alpha},& \text{if $\tilde{\Cc}$ is MSR and } 1\leq j\leq \tilde{k} \\
    \tilde{\alpha}-(j-1)\tilde{\beta},              & \text{if $\tilde{\Cc}$ is MBR and } 1\leq j\leq \tilde{k} \\
    0,              & \text{if $\tilde{\Cc}$ is MSR/MBR and } \tilde{k}+1\leq j\leq \tilde{n} \\
\end{cases}
\end{equation} 
Note that because of projective plane property, connecting $b-1$ blocks would result in getting $k_cr$ evaluations for $v-\kappa$ points and $k_c (r-1)$ evaluations for $\kappa$ points. Hence DC can decode the message by using an outer Gabidulin code if
\begin{equation} 
\sum_{t=1}^{v-\kappa}\sum_{j=1}^{k_c r}\tilde{a}_{j} + \sum_{t=1}^{\kappa}\sum_{j=1}^{k_c (r-1)}\tilde{a}_{j}\geq K.
\end{equation} 
%Similarly, for $b_{c}=b-2$, DC will get $k_cr$ evaluations for $v-(2\kappa-1)$ points, $k_c(r-1)$ evaluations for $2\kappa-2$ points and $k_c(r-2)$ evaluations for 1 point. Therefore, decoding at DC is possible if 
Similarly, for $b_{c}=b-2$, decoding at DC is possible if 
\begin{equation} 
\sum_{t=1}^{v-(2\kappa-1)}\sum_{j=1}^{k_c r}\tilde{a}_{j} + \sum_{t=1}^{2\kappa-2}\sum_{j=1}^{k_c (r-1)}\tilde{a}_{j}+\sum_{j=1}^{k_c (r-2)}\tilde{a}_{j}\geq K.
\end{equation}
With such an approach, for $b_{c}\leq b-3$ there are multiple collection possibilities for DC. For example, by connecting $b-3$ blocks DC can observe either a) $k_cr$ evaluations for $v-(3\kappa-2)$, $k_c(r-1)$ evaluations for $3(\kappa-1)$ points and $k_c(r-3)$ evaluations for 1 point, or b) $k_cr$ evaluations for $v-(3\kappa-3)$, $k_c(r-1)$ evaluations for $3(\kappa-2)$ points and $k_c(r-2)$ evaluations for 3 points. Therefore, we need to ensure that minimum rank accumulations of all cases is at least $K$.

\subsection{Concluding remarks}

We introduced the framework of block failure resilient (BFR) codes that can recover data stored in the system from a subset of available blocks with a load balancing property. Repairability is studied, file size bounds are derived, BFR-MSR and BFR-MBR points are characterized, explicit code constructions for a wide set of parameters are provided.

%%%%%%%%%%%%%%%%%%%%%%%%%%%%%%%%%%%%%%%%%%%%%%%%%%%%%%%%%%%%%%%%%%%%%%%%%%%%%%
%%%%%%%%%%%%%%%%%%%%%%%%%%%%%%%%%%%%%%%%%%%%%%%%%%%%%%%%%%%%%%%%%%%%%%%%%%%%%%%

%% REFERENCES

% Generated by IEEEtran.bst, version: 1.13 (2008/09/30)


\begin{thebibliography}{10}
\providecommand{\url}[1]{#1}
\csname url@samestyle\endcsname
\providecommand{\newblock}{\relax}
\providecommand{\bibinfo}[2]{#2}
\providecommand{\BIBentrySTDinterwordspacing}{\spaceskip=0pt\relax}
\providecommand{\BIBentryALTinterwordstretchfactor}{4}
\providecommand{\BIBentryALTinterwordspacing}{\spaceskip=\fontdimen2\font plus
\BIBentryALTinterwordstretchfactor\fontdimen3\font minus
  \fontdimen4\font\relax}
\providecommand{\BIBforeignlanguage}[2]{{%
\expandafter\ifx\csname l@#1\endcsname\relax
\typeout{** WARNING: IEEEtran.bst: No hyphenation pattern has been}%
\typeout{** loaded for the language `#1'. Using the pattern for}%
\typeout{** the default language instead.}%
\else
\language=\csname l@#1\endcsname
\fi
#2}}
\providecommand{\BIBdecl}{\relax}
\BIBdecl

\bibitem{Dimakis:Network10}
A.~G.~Dimakis, P.~B.~Godfrey, Y.~Wu, M.~J.~Wainwright, and K.~Ramchandran, ``{N}etwork
  coding for distributed storage systems,'' \emph{{IEEE} Trans. Inf. Theory},
  vol.~56, no.~9, pp. 4539--4551, Sep. 2010.

\bibitem{Tamo:Zigzag13}
I.~Tamo, Z.~Wang, and J.~Bruck, ``{Z}igzag codes: {MDS} array codes with
  optimal rebuilding,'' \emph{{IEEE} Trans. Inf. Theory}, vol.~59, no.~3, pp.
  1597--1616, Mar. 2013.

\bibitem{Rashmi:Optimal11}
K.~V.~Rashmi, N.~B.~Shah, and P.~V.~Kumar, ``{O}ptimal exact-regenerating codes for
  distributed storage at the {MSR} and {MBR} points via a product-matrix
  construction,'' \emph{{IEEE} Trans. Inf. Theory}, vol.~57, no.~8, pp.
  5227--5239, Aug. 2011.

\bibitem{Dimakis:Survey11}
A.~G.~Dimakis, K.~Ramchandran, Y.~Wu, and C.~Suh, ``{A} survey on network codes
  for distributed storage,'' \emph{Proc. {IEEE}}, vol.~99, no.~3, pp. 476--489,
  Mar. 2011.

\bibitem{Gopalan:Locality12}
P.~Gopalan, C.~Huang, H.~Simitci, and S.~Yekhanin, ``{O}n the locality of
  codeword symbols,'' \emph{{IEEE} Trans. Inf. Theory}, vol.~58, no.~11, pp.
  6925--6934, Nov. 2012.

\bibitem{Papailiopoulos:Locally12}
D.~S.~Papailiopoulos and A.~G.~Dimakis, ``{L}ocally repairable codes,'' in \emph{Proc. 2012
  IEEE International Symposium on Information Theory (ISIT 2012)}, Boston, MA,
  Jul. 2012.

\bibitem{Oggier:Self11}
F.~Oggier and A.~Datta, ``{S}elf-repairing homomorphic codes for distributed
  storage systems,'' in \emph{Proc. 2011 IEEE INFOCOM}, Shanghai, China, Apr.
  2011.

\bibitem{Rawat:Optimal14}
A.~S.~Rawat, O.~O.~Koyluoglu, N.~Silberstein, and S.~Vishwanath, ``{O}ptimal
  locally repairable and secure codes for distributed storage systems,''
  \emph{{IEEE} Trans. Inf. Theory}, vol.~60, no.~1, pp. 212--236,
  Jan. 2014.

\bibitem{Kamath:Codes12}
G.~M.~Kamath, N.~Prakash, V.~Lalitha, and P.~V.~Kumar, ``{C}odes with local
  regeneration,'' \emph{CoRR}, vol. abs/1211.1932, Nov. 2012.

\bibitem{Kamath:Explicit13}
G.~M.~Kamath, N.~Silberstein, N.~Prakash, A.~S.~Rawat, V.~Lalitha, O.~O.~Koyluoglu, P.~V.~Kumar, and S.~Vishwanath, ``{E}xplicit {MBR} all-symbol
  locality codes,'' in \emph{Proc. 2013 IEEE International Symposium on Information
  Theory (ISIT 2013)}, Istanbul, Turkey, Jul. 2013.

\bibitem{Sathiamoorthy:XORing13}
M.~Sathiamoorthy, M.~Asteris, D.~Papailiopoulos, A.~G. Dimakis, R.~Vadali,
  S.~Chen, and D.~Borthakur, ``{XOR}ing elephants: Novel erasure codes for big
  data,'' \emph{Proc. VLDB Endow.}, vol.~6, no.~5, pp. 325--336, Mar. 2013.

\bibitem{Huang:Erasure12}
C.~Huang, H.~Simitci, Y.~Xu, A.~Ogus, B.~Calder, P.~Gopalan, J.~Li, and
  S.~Yekhanin, ``{E}rasure coding in {W}indows azure storage,'' in \emph{Proc. USENIX
  Annual Technical Conference}, Boston, MA, Jun. 2012.

\bibitem{Ghemawat:Google03}
S.~Ghemawat, H.~Gobioff, and S.-T. Leung, ``{T}he {G}oogle file system,'' in
  \emph{Proc. Nineteenth ACM Symposium on Operating Systems
  Principles}, Bolton Landing, NY, Oct. 2003.

\bibitem{Ford:Availability10}
D.~Ford, F.~Labelle, F.~I. Popovici, M.~Stokely, V.-A. Truong, L.~Barroso,
  C.~Grimes, and S.~Quinlan, ``{A}vailability in globally distributed storage
  systems,'' in \emph{Proc. 9th USENIX Symposium on Operating Systems Design and
  Implementation}, Vancouver, BC, Oct. 2010.

\bibitem{Ishai:Batch04}
Y.~Ishai, E.~Kushilevitz, R.~Ostrovsky, and A.~Sahai, ``{B}atch codes and their
  applications,'' in \emph{Proc. Thirty-sixth Annual ACM Symposium
  on Theory of Computing}, Chicago, IL, Jun. 2004.

\bibitem{Gaston:realistic13}
B.~Gaston, J.~Pujol, and M.~Villanueva, ``{A} realistic distributed storage
  system: The rack model,'' \emph{CoRR}, vol. abs/1302.5657, Feb. 2013.

\bibitem{Rashmi:Enabling11}
K.~V.~Rashmi, N.~B.~Shah, and P.~V.~Kumar, ``{E}nabling node repair in any erasure
  code for distributed storage,'' in \emph{Proc. 2011 IEEE International
  Symposium on Information Theory (ISIT 2011)}, Saint Petersburg, Russia, Jul.
  2011.

\bibitem{Gabidulin:Theory85}
E.~M. Gabidulin, ``{T}heory of codes with maximum rank distance,''
  \emph{Problemy Peredachi Informatsii}, vol.~21, no.~1, pp. 3--16, 1985.

\bibitem{McWilliams:Theory77}
F.~J. McWilliams and N.~J.~A. Sloane, \emph{{T}he theory for error-correcting
  codes}.\hskip 1em plus 0.5em minus 0.4em\relax North-Holland, 1977.

\bibitem{Ho:random06}
T.~Ho, M.~Medard, R.~Koetter, D.~R.~Karger, M.~Effros, J.~Shi, and B.~Leong, ``{A}
  random linear network coding approach to multicast,'' \emph{{IEEE} Trans.
  Inf. Theory}, vol.~52, no.~10, pp. 4413--4430, Oct. 2006.

\bibitem{Stinson:Combinatorial04}
D.~R. Stinson, \emph{{C}ombinatorial designs: construction and analysis}.\hskip
  1em plus 0.5em minus 0.4em\relax Springer, 2004.

\end{thebibliography}
\end{document}